\documentclass[12pt,a4paper]{article}

\newcommand{\oa}{\omega_0}
\newcommand{\ob}{\omega_1}
\newcommand{\za}{\zeta_1}
\newcommand{\zb}{\zeta_2}
\newcommand{\xa}{\xi_1}
\newcommand{\xb}{\xi_2}
\newcommand{\sn}{\mathrm{sn}}
\newcommand{\dn}{\mathrm{dn}}
\newcommand{\oc}{\omega_{10}}
\newcommand{\me}{\mathcal{E}}
\newcommand{\ms}{\mathcal{S}}
\newcommand{\la}{\lambda}
\newcommand{\cn}{\mathrm{cn}}
\newcommand{\am}{\mathrm{am}}

\begin{document}

\begin{flushright}
{ }
\end{flushright}
\vspace{1.8cm}

\begin{center}
 \textbf{\Large Spiky Strings with Two Spins in $AdS_5$}
\end{center}
\vspace{1.6cm}
\begin{center}
 Shijong Ryang
\end{center}

\begin{center}
\textit{Department of Physics \\ Kyoto Prefectural University of Medicine
\\ Taishogun, Kyoto 603-8334 Japan}
\par
\texttt{ryang@koto.kpu-m.ac.jp}
\end{center}
\vspace{2.8cm}
\begin{abstract}
Using the reduction of the string sigma model to the 1-d Neumann
integrable model we reconstruct the closed string solution with
two equal spins in $AdS_5$ which is specified by the number of
round arcs and one winding number. From the string sigma model
itself as well as its reducion to the Neumann-Rosochatius system
we construct a spiky string solution with two unequal spins 
in $AdS_5$ whose ratio is fixed and derive its energy-spin relation.
The string configuration is characterized by the number of spikes
and two equal winding numbers associated with the two rotating
angular directions.
\end{abstract}
\vspace{3cm}
\begin{flushleft}
March, 2010
\end{flushleft}

\newpage
\section{Introduction}

The AdS/CFT correspondence \cite{JM} has more and more revealed the deep
relations between the $\mathcal{N}=4$ super Yang-Mills (SYM) theory
and the string theory in $AdS_5 \times S^5$, where various types of
classical string solutions play an important role. The energy spectrum of
certain string states matches with the spectrum of dimensions of field
theory operators in the SYM theory. There has been
a mounting evidence that the spectrum of AdS/CFT is described by
studying the multi-spin folded or circular rotating string solutions
in $AdS_5 \times S^5$  in a particular large spin limit
\cite{GKP,FT,SFT} and by analyzing the Bethe equation
for the diagonalization of the integrable spin chain in the planar SYM
theory \cite{MZ,BS}. 

Specially the energy of the one-spin folded rotating string in 
$AdS_3$ \cite{GKP,VE} describes the dimension of twist two gauge theory
operator. The large spin limit of it is related \cite{KRTT} via an
analytic continuation and an SO(2,4) transformation to the open string
solution ending on a null cusp at the boundary which determines the
planar four-point gluon amplitude at 
strong coupling in the SYM theory \cite{AM,MK,MMT,JAM,AGM}.
The quantum corrections to the folded string energy have been computed
up to two loops in the long string limit \cite{FT,KRTT,RT,BF,RRTT} 
and have been tested to match with the prediction
of the strong-coupling expansion of the solution \cite{BB} of the
integral equation for the minimal twist anomalous dimensions as
extracted from the weak-coupling all-loop asymptotic Bethe 
ansatz \cite{BES,BS}. This matching has been extended to include the
spin $J$ in $S^1 \subset S^5$ \cite{FTT,CK}.

The spiky string solution in $AdS_3$ has been constructed by using the
Nambu-Goto string action as a generalization of the one-spin folded
string solution to the case of many spikes \cite{MKR}. 
This string state has been argued to correspond to a subclass of 
higher twist operators in the SL(2) sector of gauge theory \cite{BGK}.
The angular separation between each of adjacent spikes is the same.
Using the string sigma-model action in the conformal gauge the
Kruczenski's spiky string solution with one spin has been reproduced
in \cite{JJ}. The spiky string solution in $AdS_3$ with different angular
separations has been constructed \cite{DL} by using the finite-gap
formalism for the string sigma-model action \cite{BGK,KZ}, where
each classical solution of string theory has an associated spectral 
elliptic curve which encodes the values of the higher conserved charges
of the worldsheet sigma model. A different attempt to generalize the
spiky string solution has been presented \cite{AJJ}, 
where a construction of a general class of string solution in $AdS_3$ is
based on a Pohlmeyer type reduction \cite{KP} with the sinh-Gordon model,
whose N-soliton solutions map to the dynamical N-spike AdS string
configurations.

For the spiky string in $AdS_3$ in the large spin limit the spikes
approach the boundary of $AdS_3$ and their motion can be described by a
string solution in an $AdS_5-$pp-wave metric \cite{KT}. Further various
solutions for open strings moving in the $AdS_3-$pp-wave space have been
produced such that they are dual to various Wilson loops in the field 
theory in the boundary 4-dimensional pp-wave background \cite{IKT}.
There has been an investigation of the spiky string solution in 
$AdS_3 \times S^1$ with $n$ spikes and spin $S$ in $AdS_3$ and 
spin $J$ and winding $m$ in $S^1$ \cite{RIK}. In a special large $n$ 
limit the solution can be regarded as describing a periodic-spike string
in $AdS_3-$ pp-wave$\times S^1$ background. On the other hand based on the
SL(2) asymptotic Bethe ansatz equations, the rational $(S, J)$ solution
has been constructed as the one-cut solution with non-trivial winding
\cite{NZZ} and the spiky string solution with winding $m$, spikes $n$ and
two spins $S, J$ has been described as the two-cut solution \cite{KKT}
by following the general procedure of solving integral Bethe ansatz 
equation at strong coupling \cite{FKT}. The energy of this two-cut 
solution matches with that of the spiky string solution presented 
in \cite{RIK} in a special scaling limit with large two spins and large
winding where the string touches the $AdS_5$ boundary.

The single-spin folded string in $AdS_3$ has been extended so that
the string solutions with two equal spins $S_1 = S_2$ in $AdS_5$
\cite{LK} and further with three spins $S_1 = S_2$ and $J$
in $AdS_5\times S^1$ \cite{SRY} are produced by using string sigma-model
actions in the conformal gauge, where the closed string configurations 
consisting of round arcs are characterized by the number of arcs $n$ and
the winding number $m$. The large spin behavior of the minimal energy
specified by $n = 3, \; m = 1$ for the string solution
with  two equal spins $S_1 = S_2$
in $AdS_5$ has been argued \cite{TT} to match with the strong-coupling 
prediction from the analysis \cite{FRZ} of the full asymptotic Bethe
ansatz equation \cite{BS,BES}. In the approach of \cite{AF} that the 
conformal-gauge string sigma model is reduced to the 1-d Neumann 
integrable model, the closed string solutions in $AdS_5$ are parametrized
by the three frequencies $\omega_a = (\oa, \ob, \omega_2), \oa = \kappa$
and the two integrals of motion $b_1, \; b_2$. In the special 
$\kappa = \omega_2 \ne \ob$ case an explicit 
solution with two unequal spins
$S_1 \ne S_2$ has been constructed by making suitable change of global 
coordinates for the conformal-gauge string sigma-model action itself
\cite{TT}. Further for the other special $b_1 = b_2$ case the energy-spin
relation of circular string solution with two unequal spins has been 
derived by analyzing the 1-d Neumann integrable model. 

Using the approach of \cite{ART} that the conformal-gauge string sigma
model is reduced to the integrable Neumann-Rosochatius system,
a solution which expresses a hanging string moving with two spins
in $AdS_5$ has been presented \cite{IK} and shown to correspond to
the double-helix Wilson loop describing a quark and an anti-quark moving
on an $S^3$. 

We will consider the 1-d Neumann integrable model for the $\ob = \omega_2$
case of the closed string in $AdS_5$ to reconstruct the string solution 
of round arc shape with two equal spins  $S_1 = S_2$.
In order to extend the spiky string in $AdS_3$ to in $AdS_5$
we will derive a spiky string solution with two spins 
by analyzing the conformal-gauge string sigma model itself as well as
 its reduction to the integrable Neumann-Rosochatius system.
The energy-spin relations for both string solutions in the large spin
limit will be compared.

\section{Strings of round arc shape with two equal spins}

We consider a rigid rotating two-spin string in $AdS_5$ based on 
the reduction to the 1-d Neumann integrable model 
\cite{AF}. For the closed bosonic string in the
conformal gauge in $AdS_5$ space parametrized by global coordinates
\begin{equation}
ds^2 = - \cosh^2\rho dt^2 + d\rho^2 + \sinh^2\rho ( d\theta^2 + 
\sin^2\theta d\phi_1^2 + \cos^2\theta d\phi_2^2 )
\label{gl}\end{equation}
we make the following ansatz
\begin{equation} 
t = \kappa\tau, \;\; \rho = \rho(\sigma), \;\; \theta = \theta(\sigma), 
\;\; \phi_1 = \ob\tau, \;\; \phi_2 = \omega_2\tau.
\end{equation}
In terms of the three complex embedding coordinates $X_a, \; a=0,1,2$
subject to a constraint
\begin{equation} 
-1 = |X_1|^2 + |X_2|^2 - |X_3|^2 = \sum_a \eta_a X_a \bar{X}_a 
\label{ex}\end{equation}
with $\eta_0 = -1, \eta_1 = \eta_2 = 1$, the $AdS_5$ metric (\ref{gl}) is
rewritten by
\begin{equation}
ds^2 = \sum_a \eta_a dX_a d\bar{X}_a,
\label{ds}\end{equation}
where 
\begin{eqnarray}
X_0 &=& r_0e^{it}, \hspace{1cm} X_1 = r_1e^{i\phi_1}, \hspace{1cm} 
X_2 = r_2e^{i\phi_2}, \nonumber \\
r_0 &=& \cosh\rho,  \hspace{1cm} r_1 = \sinh\rho \sin \theta, 
\hspace{1cm} r_2 = \sinh\rho \cos\theta.
\end{eqnarray}

Following the approach of ref. \cite{AF,TT} we use the two independent
coordinates $\zeta_i, \; i = 1, 2$ related to $r_a$ as
\begin{equation}
r_a^2 = - \eta_a \frac{(\za - \omega_a^2)(\zb - \omega_a^2)}
{\prod_{b \ne a} \omega_{ab}^2 },
\label{rg}\end{equation}
where $\omega_a = (\oa, \ob, \omega_2), \oa = \kappa$ and 
$\omega_{ab}^2 \equiv \omega_a^2 - \omega_b^2$. In terms of the 
coordinates $\zeta_i(\sigma)$ the string equations of motion read
\begin{equation}
{\za'}^2 = -4 \frac{P_5(\za)}{(\zb - \za)^2}, \hspace{1cm}
{\zb'}^2 = -4 \frac{P_5(\zb)}{(\zb - \za)^2},
\label{st}\end{equation}
where a quintic polynomial is expressed as
\begin{equation}
P_5(\zeta) = (\zeta - \oa^2)(\zeta - \ob^2)(\zeta - \omega_2^2) 
(\zeta - b_1)(\zeta - b_2)
\label{pf}\end{equation}
with two constants of motion $b_1, \;b_2$ obeying 
\begin{equation}
b_1 + b_2 = \oa^2 + \ob^2 + \omega_2^2.
\label{bb}\end{equation}
The $b_1, \;b_2$ parameters are associated with the integrals of motion of
the 1-d Neumann integrable model. The two-spin solution of a circular 
shape is specified by the following range
\begin{equation}
\oa^2 \le \omega_2^2 \le \za \le \ob^2 \le b_1 \le \zb \le b_2.
\label{ra}\end{equation}

We consider the $\ob = \omega_2$ case corresponding to the rotating string
with two equal spins in $AdS_5$. For the equations in (\ref{st})
we make the change of variables
\begin{equation}
\za = \ob^2 - \omega_{12}^2 \xa, \hspace{1cm} \zb = b_2 - b_{21}\xb
\end{equation}
with $b_{21} \equiv b_2 - b_1$ and then take the limit 
$\omega_2 \rightarrow \ob$ to obtain the equations of motion for 
$\xa, \; \xb$
\begin{eqnarray}
{\xa'}^2 &=& 4h (1 - u)\left( 1 - \frac{t}{u} \right) \frac{\xa(1 - \xa)}
{(1 - u \xb)^2}, \label{fir} \\
{\xb'}^2 &=& 4h \xb(1 - \xb)(1 - t \xb), 
\label{sec}\end{eqnarray}
where the parameters $h, \; u$ and $t$ are given by
\begin{equation}
h = b_2 - \oa^2 > 0, \hspace{1cm} u = \frac{b_{21}}{b_2 - \ob^2} > 0,
\hspace{1cm} t = \frac{b_{21}}{b_2 - \oa^2} > 0.
\label{hu}\end{equation}
The string coordinates $r_a$ in the limit $\omega_2 \rightarrow \ob$ 
are expressed in terms of $\xa, \; \xb$ as
\begin{eqnarray}
r_0^2 &=& \frac{b_2 - \oa^2}{\oc^2}(1 - t \xb), \nonumber \\
r_1^2 &=& \frac{b_2 - \ob^2}{\oc^2}\xa(1 - u \xb), \hspace{1cm}
r_2^2 = \frac{b_2 - \ob^2}{\oc^2}(1 - \xa)(1 - u \xb).
\label{rb}\end{eqnarray}

The equation (\ref{sec}) is integrated to be expressed in terms of the 
Jacobi elliptic function sn and the complete elliptic integral 
of the first kind $K$
\begin{equation}
\xb(\sigma) = \sn^2 ( K(t) - \sqrt{h}\sigma, t),
\label{sn}\end{equation}
where we choose that $\xb(\sigma)$ decreases in the interval
$0 < \sigma < \pi/2$ and fix an integration constant as 
$\xb(0)= 1$ so that $\xb(\sigma_0) = 0$ with 
$\sigma_0 \equiv K(t)/\sqrt{h}$. The substitution of the solution 
(\ref{sn}) into $r_0^2$ in (\ref{rb}) yields
\begin{equation}
\cosh^2\rho = \cosh^2\rho_1 \dn^2(K - \sqrt{h}\sigma,t)
\label{dn}\end{equation}
with $\cosh^2\rho_1 = (b_2 - \oa^2)/\oc^2$ so that 
$\cosh^2\rho(\sigma_0) = \cosh^2\rho_1$. We define 
$\cosh^2\rho_0 = (b_1 - \oa^2)/\oc^2$ to have  $\cosh^2\rho(0) 
= \cosh^2\rho_0$. The interval $b_1 \le \zb \le b_2$ in (\ref{ra}) implies
$\rho_0 \le \rho_1$. We see that $\rho(\sigma)$ starts off at its minimum
$\rho_0$ at $\sigma = 0$, then increases to its maximum $\rho_1$ at 
$\sigma = \sigma_0$ in one segment. By gluing together $2n_1$ such
segments that are $n_1$ arcs to impose the periodicity condition
$\rho(\sigma + 2\pi) = \rho(\sigma)$, we have 
\begin{equation}
\sqrt{h} \frac{2\pi}{n_1} = 2K(t)
\label{hk}\end{equation}
because the period of dn is $2K$.
The expression (\ref{dn}) is rewritten by
\begin{equation}
\cosh^2\rho = \frac{\cosh^2\rho_0}{\dn^2(\sqrt{h}\sigma, t)},
\end{equation}
where $t$ and $\sqrt{h}$ are described in terms of $\cosh\rho_1, \;
\cosh\rho_0$ as
\begin{equation}
t = \frac{\cosh^2\rho_1 - \cosh^2\rho_0}{\cosh^2\rho_1}, \hspace{1cm}
\sqrt{h} = \sqrt{\oc^2}\cosh\rho_1.
\label{ht}\end{equation}

The other equation (\ref{fir}) is also integrated to be 
\begin{equation}
\xa = \sin^2 \left( \sqrt{h(1 - u)\left(1 - \frac{t}{u}\right)} 
\int_{\sigma_0}^{\sigma} \frac{d\sigma}{1 - u \xb} \right).
\end{equation}
Since the string configuration (\ref{rb}) implies 
\begin{equation}
\tan^2\theta = \frac{r_1^2}{r_2^2} = \frac{\xa}{1 - \xa}
\label{ta}\end{equation}
we have
\begin{equation}
\theta(\sigma) = \sqrt{h(1 - u)\left(1 - \frac{t}{u}\right)}
\int_{\sigma_0}^{\sigma} \frac{d\sigma}{1 - u \xb},
\label{th}\end{equation}
where an integration constant is chosen such that $\theta(\sigma =
\sigma_0) = 0\; (\mathrm{i.e.} \; 
\theta(\rho = \rho_1) = 0)$ for convenience.
Substituting the solution (\ref{sn}) into (\ref{th}) 
and integrating we derive
\begin{equation}
\theta(\sigma) = - \sqrt{ \frac{\cosh^2\rho_0 - 1}
{\cosh^2\rho_1(\cosh^2\rho_1 - 1)} }
\Pi(\am(K - \sqrt{h}\sigma), u, t)
\label{ths}\end{equation}
with $u = (\cosh^2\rho_1 - \cosh^2\rho_0)/(\cosh^2\rho_1 - 1)$,
where $\Pi (x,u,t)$ is the incomplete of elliptic integral of the third 
kind. From the periodicity condition $\theta(2\pi) - \theta(0) = 2\pi n_2$
for the angular coordinate an integer winding number $n_2$ is specified by
\begin{equation}
2\pi n_2 = 2 n_1( \theta(\sigma_0) - \theta(0) ) \equiv 2 n_1 
\Delta\theta, \hspace{1cm} n_2 = 1, 2, 3, \cdots,
\end{equation}
which together with (\ref{ths}) yields
\begin{equation}
\frac{n_2}{n_1} \pi = \sqrt{ \frac{\cosh^2\rho_0 - 1}{\cosh^2\rho_1
(\cosh^2\rho_1 - 1)} }\Pi(u, t) \equiv \Delta \theta.
\end{equation}

For the $\ob \ne \omega_2$ case we write down the energy and two spins
of rotating string
\begin{eqnarray}
\me &=& \oa \int_0^{2\pi}\frac{d\sigma}{2\pi} r_0^2 = 
\frac{\oa(b_2 - \oa^2)}{\omega_{20}^2}\int_0^{2\pi}\frac{d\sigma}{2\pi} 
(1 - t\xb) \left( 1 - \frac{\omega_{12}^2}{\oc^2}\xa \right), \nonumber \\
\ms_1 &=& \ob \int_0^{2\pi}\frac{d\sigma}{2\pi} r_1^2 = 
\frac{\ob(b_2 - \ob^2)}{\oc^2}\int_0^{2\pi}\frac{d\sigma}{2\pi} 
\xa \left( 1 - \frac{b_{21}}{b_2 - \ob^2}\xb \right), \nonumber \\
\ms_2 &=& \omega_2 \int_0^{2\pi}\frac{d\sigma}{2\pi} r_2^2 = 
\frac{\omega_2(b_2 - \omega_2^2)}{\omega_{20}^2}\int_0^{2\pi}
\frac{d\sigma}{2\pi} (1 - \xa) \left( 1 - \frac{b_{21}}{b_2 - \omega_2^2}
\xb \right).
\end{eqnarray}
For the $\ob = \omega_2$ case the energy and total spin 
 are simplified to be
\begin{eqnarray}
\me &=& \frac{\oa(b_2 - \oa^2)}{\omega_{10}^2}\int_0^{2\pi}
\frac{d\sigma}{2\pi}(1 - t\xb), \nonumber \\
\ms &=& \ms_1 + \ms_2 = 2\ms_1
= \frac{\ob(b_2 - \ob^2)}{\oc^2}\int_0^{2\pi}\frac{d\sigma}{2\pi} 
 ( 1 - u\xb ).
\end{eqnarray}
Here from (\ref{hu}) we obtain $b_1, \;b_2$ as
\begin{equation}
b_1 = \frac{t(1 - u)\oa^2 + u(t - 1)\ob^2}{t - u}, \hspace{1cm}
b_2 = \frac{t\oa^2 - u\ob^2}{t - u},
\end{equation}
which combine with (\ref{bb}) to give
\begin{equation}
\frac{\oa^2}{\ob^2} = \frac{t(2 - u)}{u + t - ut}.
\label{tu}\end{equation}
Gathering (\ref{hk}), (\ref{ht}) and (\ref{tu}) together we have
\begin{equation}
\ob = \frac{ \sqrt{\cosh^2\rho_1 + \cosh^2\rho_0 - 1} }{\cosh\rho_1}
\frac{n_1K(t)}{\pi}.
\label{on}\end{equation}
Therefore the two constants of motion $b_1, b_2$ can be expressed in terms
of $\rho_0, \rho_1$. We use (\ref{tu}) and the explicit expression 
$\xb$ of (\ref{sn}) to derive
\begin{eqnarray}
\me &=& \frac{n_1}{\pi} \frac{\oa}{\oc^2}\sqrt{h}E(t) 
= \frac{n_1}{\pi}\sqrt{ut(2 - u)}\frac{E(t)}{u - t}, \nonumber \\
\ms &=& \frac{n_1}{\pi}\frac{\ob}{\sqrt{h}}\left( \frac{uE(t)}{u - t} 
- K(t) \right) = \frac{n_1}{\pi}\sqrt{ \frac{u + t - ut}{u} }
\left( \frac{uE(t)}{u - t} - K(t) \right), 
\end{eqnarray}
where $E$ is the complete elliptic integral of the second kind.

Thus we have reproduced the same expressions for $\me, \ms, \Delta\theta$
 as in \cite{LK,SRY,TT}
by means of the reduction to the 1-d Neumann integrable model.
This equal two-spin string configuration consists of $n_1$ round arcs
with equal angular separation $2\Delta\theta$.
In the large spin limit $u \approx t \approx 1$ in $t \le u$
which is associated with $\oa \le \ob$ in (\ref{tu}), these expressions
yield the following energy-spin relation
\begin{equation}
E - S \approx \frac{n_1\sqrt{\la}}{2\pi} \left( \ln \frac{16\pi S}{n_1
\sqrt{\la}} - 1 + 2\ln \sin \frac{n_2\pi}{n_1} \right),
\label{ns}\end{equation}
where $\sqrt{\la}$ is the string tension and
the subleading contributions are presented in \cite{TT}.

\section{Spiky strings with two spins}

We consider a spiky string with two spins in $AdS_5$ in the global 
coordinates of (\ref{gl}) using the string sigma-model action
in the conformal gauge. We choose equal two frequencies $\ob = \omega_2
= \omega$ and parametrize the closed string as
\begin{eqnarray}
t &=& \oa \tau + \mu_0(\sigma), \hspace{1cm} \phi_1 = \omega \tau + 
\mu_1(\sigma), \hspace{1cm} \phi_2 = \omega \tau + \mu_2(\sigma),
\nonumber \\
\rho &=& \rho(\sigma), \hspace{1cm} \theta = \theta(\sigma).
\label{pat}\end{eqnarray}
The equations of motion for $t, \phi_1$ and $\phi_2$ lead to
\begin{equation}
\mu'_0 = - \frac{C_0}{\cosh^2\rho}, \hspace{1cm} \mu'_1 =  \frac{C_1}
{\sinh^2\rho\sin^2\theta}, \hspace{1cm} \mu'_2 =  \frac{C_2}
{\sinh^2\rho\cos^2\theta}
\end{equation}
with three integration constants $C_0, C_1, C_2$. The Virasoro 
constraints read
\begin{eqnarray}
\oa C_0 &+& \omega( C_1 + C_2 ) = 0,
\label{cc} \\
{\rho'}^2 &+& \sinh^2\rho {\theta'}^2 - \cosh^2\rho \left( \oa^2 +
\frac{C_0^2}{\cosh^4\rho} \right) \nonumber \\
&+& \sinh^2\rho \left( \omega^2 + \frac{C_1^2}{\sinh^4\rho\sin^2\theta} 
+ \frac{C_2^2}{\sinh^4\rho\cos^2\theta} \right) = 0.
\label{hr}\end{eqnarray}
In view of the equation of motion for $\theta$
\begin{equation}
 \frac{\partial}{\partial \sigma}(\sinh^2\rho \theta') -
\frac{\sin\theta \cos\theta}{\sinh^2\rho} \left(\frac{C_1^2}{\sin^4\theta}
-  \frac{C_2^2}{\cos^4\theta} \right) = 0
\end{equation}
we obtain a solution $\theta = \theta_0$ with
\begin{equation}
\tan^2\theta_0 = \frac{C_1}{C_2},
\label{pa}\end{equation}
whose special case is $\theta_0 = \pi/4, C_1 = C_2$.
Then the Virasoro constraint (\ref{hr}) is expressed 
in a factorized form as
\begin{equation}
{\rho'}^2 = \frac{(\oa^2\cosh^2\rho - \omega^2\sinh^2\rho)
\left(\sinh^22\rho - \frac{4C_0^2}{\omega^2}\right)}{\sinh^22\rho},
\end{equation}
where we use two relations $C_1 = -\oa C_0\sin^2\theta_0/\omega, 
C_2 = -\oa C_0\cos^2\theta_0/\omega$ derived 
from (\ref{cc}) and (\ref{pa}).

Alternatively we analyze this spiky string by using the reduction to the
integrable Neumann-Rosochatius system \cite{ART,IK}.
The closed string is parametrized in terms of the embedding coordinates
$X_a$ in (\ref{ex}), (\ref{ds}) as
\begin{equation}
X_a = r_a e^{i(\mu_a(\sigma) + \omega_a\tau )}, \hspace{1cm}
a = 0, 1, 2
\label{xm}\end{equation}
with $\ob = \omega_2 = \omega$ and $\sum_a \eta_a r_a^2 = -1$.
Following the prescription for the string in $AdS_5$ presented in
\cite{IK}, the functions $\mu_a(\sigma)$ are characterized by
\begin{equation}
\mu'_a = \frac{\eta_a C_a}{r_a^2}.
\label{mu}\end{equation}
One of the Virasoro constraints yields
\begin{equation}
\sum_a \omega_a C_a = 0,
\label{ca}\end{equation}
which is the same as (\ref{cc}). This system is described by the two 
unconstrained coordinates $\zeta_i(\sigma), \; i =1, 2$ in (\ref{rg})
and solved by using the Hamiltonian-Jacobi method. The string equations of
motion are also expressed by (\ref{st}) but the quintic polynomial
$P_5(\zeta)$ is given by
\begin{equation}
P_5(\zeta) = -\prod_a (\zeta - \omega_a^2) \left( V - 
\sum_a \prod_{b\ne a}(\omega_a^2 - \omega_b^2)
 \frac{C_a^2}{\zeta - \omega_a^2} + \zeta \sum_a
 \omega_a^2 - \zeta^2 \right),
\end{equation}
which satisfies $P_5(\zeta_i) < 0$ and 
is compared with (\ref{pf}). The parameter $V$ is a constant of 
motion and $C_a$  are the conserved momenta canonically conjugate to
$\mu_a$ in the integrable Neumann-Rosochatius sistem.

From $r_a^2 > 0$ in (\ref{rg}) the relevant parameters 
take the following range
\begin{equation}
\oa^2 \le \omega_2^2 \le \za \le \ob^2 \le \zb
\end{equation}
in the same way as (\ref{ra}). We take the limit $\omega_2 \rightarrow 
\ob$ after the change of variable $\za = \ob^2 - \omega_{12}^2 \xa$.
The equation of motion for $\za$ is described in terms of $\xa$ as
\begin{equation}
{\xa'}^2 = -\frac{4\xa(1 - \xa)\oc^2}{(\zb - \ob^2)^2} \left( V +
\left( \frac{C_1^2}{\xa} + \frac{C_2^2}{1 - \xa} - C_0^2 \right)\oc^2
+ ( \oa^2 + \ob^2 )\ob^2  \right).
\label{vvc}\end{equation}
In order to find a solution with constant $\xa$ using (\ref{ta}) and 
(\ref{ca}) we express the relevant terms in (\ref{vvc}) as
\begin{equation}
\frac{C_1^2}{\xa} + \frac{C_2^2}{1 - \xa} = \left( \frac{\oa C_0}{\ob} 
\right)^2 \frac{ 1 + \tan^2\theta \left(\frac{C_2}{C_1}\right)^2}
{\sin^2\theta \left( 1 +  \frac{C_2}{C_1} \right)^2 },
\end{equation}
which becomes $(\oa C_0/\ob)^2$ if we choose $\theta$ to be
the same as (\ref{pa}). The string configuration with fixed
$\theta_0$ can be a solution when the following condition is satisfied
\begin{equation}
\ob^2 V + \ob^4(\oa^2 + \ob^2 ) = C_0^2(\oc^2)^2.
\label{vc}\end{equation}

In the equation of motion for $\zb$ the $P_5(\zb)$ is expressed 
owing to (\ref{vc}) as
$P_5(\zb) = (\zb - \ob^2)(\zb - \omega_2^2)f(\zb)$ with
\begin{equation}
f(\zb) = \zb^3 - 2(\oa^2 + \ob^2)\zb^2 + ( \oa^2(\oa^2 + 2\ob^2) - V )\zb
+ (\oa^2 + \ob^2)( V + \ob^4),
\end{equation}
which has three roots
\begin{eqnarray}
\la_1 &=& \frac{1}{2} \left( \oa^2 + \ob^2 - \sqrt{(\oa^2 + \ob^2)^2 + 
4( V + \ob^4)} \right), \nonumber \\
\la_2 &=& \frac{1}{2} \left( \oa^2 + \ob^2 + \sqrt{(\oa^2 + \ob^2)^2 + 
4( V + \ob^4)} \right), \nonumber \\
\la_3 &=& \oa^2 + \ob^2,
\label{ls}\end{eqnarray}
so that $f(\zb) = (\zb - \la_1)(\zb - \la_2)(\zb - \la_3)$.
The $\la_1, \la_2$ are real since the square root term is given by
$\oc^2\sqrt{ 1 + 4(C_0/\ob)^2 }$. The condition (\ref{vc}) implies
$V + \ob^2(\oa^2 + \ob^2) > 0$ which further leads to 
$f(\oa^2) = \ob^2(V + \oa^2\ob^2 + \ob^4)> 0,
f(\ob^2) = \oa^2(V + \oa^2\ob^2 + \ob^4)> 0$. We consider the case that 
the parameter $V$ changes in 
\begin{equation}
V \le - \ob^4
\label{vi}\end{equation}
such that $\la_1 < \la_2 < \la_3$. Therefore taking account of 
$P_5(\zb) < 0$, that is $f(\zb) < 0$ and 
\begin{equation}
 \la_1 + \la_2 = \oa^2 + \ob^2
\label{tlw}\end{equation}
we have the following parameter region
\begin{equation}
\la_1 \le \oa^2 \le \ob^2 \le \la_2 \le \zb \le \la_3.
\label{lz}\end{equation}
The parametrization of $\zb$ as $\zb = \la_3 - \la_{32}\xb$ with
$\la_{32} \equiv \la_3 - \la_2$ makes the equation of motion for $\zb$ 
change into
\begin{equation}
{\xb'}^2 = 4h \xb(1 - \xb)(1 - t \xb) 
\end{equation}
with $h = \la_{31} \equiv \la_3 - \la_1 >0, \; t = \la_{32}/\la_{31}> 0$.
This equation also gives a solution
\begin{equation}
\xb(\sigma) = \sn^2 ( K(t) - \sqrt{h}\sigma, t).
\label{sk}\end{equation}
The string coordinates $r_a$ (\ref{rg}) in the limit 
$\omega_2 \rightarrow \ob$ are
expressed in terms of $\xa = \sin^2\theta_0$ and $\xb(\sigma)$ as
\begin{eqnarray}
r_0^2 &=& \frac{\la_3 - \oa^2}{\oc^2}( 1 - u_0\xb ), \nonumber \\
r_1^2 &=& \frac{\la_3 - \ob^2}{\oc^2}\xa ( 1 - u_1\xb ), \hspace{1cm}
r_2^2 = \frac{\la_3 - \ob^2}{\oc^2}(1 - \xa) ( 1 - u_1\xb )
\label{la}\end{eqnarray}
with $u_0 = \la_{32}/(\la_3 - \oa^2)> 0, \; 
u_1 = \la_{32}/(\la_3 - \ob^2)> 0$. 

From (\ref{la}) with (\ref{sk}) the AdS radial coordinate $\rho$ is 
determined by
\begin{eqnarray}
\cosh^2\rho &=& \frac{\la_3 - \oa^2 - \la_{32}\sn^2(K - \sqrt{h}\sigma,t)}
{\oc^2} \nonumber \\
&=& \frac{\cosh^2\rho_1\dn^2(\sqrt{h}\sigma,t) - (\cosh^2\rho_1 
- \cosh^2\rho_0)\cn^2(\sqrt{h}\sigma,t)}{\dn^2(\sqrt{h}\sigma,t)},
\end{eqnarray}
which is also represented by
\begin{equation}
\cosh2\rho = \cosh2\rho_1\cn^2(K - \sqrt{h}\sigma,t) + 
\cosh2\rho_0\sn^2(K - \sqrt{h}\sigma,t),
\end{equation}
where 
\begin{equation}
\cosh^2\rho_1 = \frac{\la_3 - \oa^2}{\oc^2} = \frac{\ob^2}{\oc^2},
\hspace{1cm} \cosh^2\rho_0 = \frac{\la_2 - \oa^2}{\oc^2}.
\label{rh}\end{equation}
In one segment the radial coordinate $\rho$ starts off  at its minimum
$\rho_0$ at $\sigma = 0$ and increases to its maximum $\rho_1$ 
at $\sigma = \sigma_0 \equiv K(t)/\sqrt{h}$. The gluing of $2n_1$
segments to form a closed string yields a condition
\begin{equation}
\sqrt{h} \pi = n_1 K(t),
\label{nk}\end{equation}
where $t$ is expressed  through a relation (\ref{tlw}) as
\begin{equation}
t = \frac{\cosh^2\rho_1 - \cosh^2\rho_0}{\cosh^2\rho_1 + \cosh^2\rho_0 -1}
= \frac{\cosh2\rho_1 - \cosh2\rho_0}{\cosh2\rho_1 + \cosh2\rho_0},
\end{equation}
which is compared with the expression of $t$ in (\ref{ht}).

From the explicit solution (\ref{sk}) the equations (\ref{mu}) for 
$\mu_a(\sigma)$ are integrated to be
\begin{eqnarray}
\mu_0 &=& -\frac{C_0\oc^2}{\la_3 - \oa^2}\int_{\sigma_0}^{\sigma} 
\frac{d\sigma}{1 - u_0\xb} = \frac{C_0}{\cosh^2\rho_1\sqrt{h} }
\Pi(\am(K - \sqrt{h}\sigma),u_0,t),  \nonumber \\
\mu_1 &=& \frac{C_1\oc^2}{\sin^2\theta_0(\la_3 - \ob^2)}
\int_{\sigma_0}^{\sigma} \frac{d\sigma}{1 - u_1\xb} = 
-\frac{C_1}{\sin^2\theta_0\sinh^2\rho_1\sqrt{h} }
\Pi(\am(K - \sqrt{h}\sigma),u_1,t),  \nonumber \\
\mu_2 &=& \frac{C_2\oc^2}{\cos^2\theta_0(\la_3 - \ob^2)}
\int_{\sigma_0}^{\sigma} \frac{d\sigma}{1 - u_1\xb} = 
-\frac{C_2}{\cos^2\theta_0\sinh^2\rho_1\sqrt{h} }
\Pi(\am(K - \sqrt{h}\sigma),u_1,t),
\end{eqnarray}
where the integration constants are chosen such that 
$\mu_a(\sigma_0) = 0$. The elimination of $\la_2$ from (\ref{ls}) and 
(\ref{rh}) leads to
\begin{equation}
C_0^2 = \frac{1}{4} \ob^2\sinh^22\rho_0.
\label{co}\end{equation}
We choose minus sign as $C_0 = -\ob\sinh2\rho_0/2$, 
which yields
\begin{equation}
C_1 = \frac{1}{2} \oa\sin^2\theta_0 \sinh2\rho_0, \hspace{1cm}
C_2 = \frac{1}{2} \oa\cos^2\theta_0 \sinh2\rho_0
\label{tc}\end{equation}
owing to (\ref{ca}) and (\ref{pa}). The expression (\ref{co}) together 
with (\ref{vc}) reads
\begin{equation}
\frac{V + \ob^4}{\ob^4} = \frac{\sinh^2\rho_0 \cosh^2\rho_0 - 
\sinh^2\rho_1 \cosh^2\rho_1}{\cosh^4\rho_1},
\end{equation}
which indeed satisfies (\ref{vi}). 
We use $\ob/\oa = \coth\rho_1$ in (\ref{rh})
to express $h$ as
\begin{equation}
h = \frac{\cosh^2\rho_1 + \cosh^2\rho_0 - 1}{\sinh^2\rho_1}\oa^2
\label{hh}\end{equation}
and obtain 
\begin{eqnarray}
\mu_0 &=& - \frac{\sinh2\rho_0}{\sqrt{2}\cosh\rho_1
\sqrt{\cosh2\rho_1 + \cosh2\rho_0} }\Pi(\am(K-\sqrt{h}\sigma),u_0,t), 
\nonumber \\
\mu_1 &=&  \mu_2 = - \frac{\sinh2\rho_0}{\sqrt{2}\sinh\rho_1
\sqrt{\cosh2\rho_1 + \cosh2\rho_0} }\Pi(\am(K-\sqrt{h}\sigma),u_1,t), 
\end{eqnarray}
where $u_0$ and $u_1$ defined in (\ref{la}) are described by 
\begin{equation}
u_0 = \frac{ \cosh2\rho_1 - \cosh2\rho_0}{\cosh2\rho_1 + 1}, \hspace{1cm}
u_1 = \frac{ \cosh2\rho_1 - \cosh2\rho_0}{\cosh2\rho_1 - 1}
\end{equation}
and the equality $\mu_1 =  \mu_2$ is due to (\ref{tc}). The condition 
(\ref{nk}) together with (\ref{hh}) implies
\begin{equation}
\ob = \frac{\cosh\rho_1}{\sqrt{ \cosh^2\rho_1 + \cosh^2\rho_0 - 1} }
\frac{n_1K(t)}{\pi},
\label{kp}\end{equation}
which is compared to (\ref{on}).

To demand that the string is closed at constant global time $t$ we 
substitute $\tau = (t - \mu_0(\sigma))/\oa$ into the ansatz for 
$\phi_1, \phi_2$ in (\ref{xm}) and (\ref{pat}) to obtain
\begin{equation}
\phi_i(t,\sigma) = \frac{\omega_i}{\oa}t + \mu_i(\sigma) - 
\frac{\omega_i}{\oa}\mu_0(\sigma), \hspace{1cm} i = 1, 2.
\end{equation}
Introducing two integer winding numbers $m_i$ for the angular coordinates
$\phi_i$ we have the following closeness conditions
\begin{equation}
2\pi m_i = \delta\phi_i = \delta\mu_i - \frac{\omega_i}{\oa}\delta\mu_0,
\end{equation}
where $\delta\phi_i = \phi_i(t,2\pi) - \phi_i(t,0), \; 
\delta\mu_a = \mu_a(2\pi) - \mu_a(0)$. Owing to $\ob = \omega_2, 
\mu_1 = \mu_2$ they are expressed in terms of the same winding number
$m_1 = m_2 = m$ in a single form as
\begin{eqnarray}
2\pi m &=& 2n_1\Delta\phi, \nonumber \\
\Delta\phi &=& \frac{\sinh2\rho_0}{\sqrt{2}\sinh\rho_1
\sqrt{\cosh2\rho_1 + \cosh2\rho_0}} (\Pi(u_1,t) - \Pi(u_0,t)).
\label{da}\end{eqnarray}
Thus $m$ is described by two parameters $\rho_0, \rho_1$ for fixed $n_1$
and the parameter $\theta_0$ does not appear.

From (\ref{la}) and (\ref{sk}) the energy $\me$ and total spin
$\ms = \ms_1 + \ms_2$ can be computed to be
\begin{eqnarray}
\me &=& \frac{\oa(\la_3 - \oa^2)}{\oc^2}\int_0^{2\pi}\frac{d\sigma}{2\pi}
(1 - u_0\xb) = \frac{n_1}{\pi} \frac{\oa}{\oc^2\sqrt{h}}( \la_{31}E
- (\oa^2 - \la_1)K ), \nonumber \\
\ms &=& \frac{\ob(\la_3 - \ob^2)}{\oc^2}\int_0^{2\pi}\frac{d\sigma}{2\pi}
(1 - u_1\xb) = \frac{n_1}{\pi} \frac{\ob}{\oc^2\sqrt{h}}( \la_{31}E
- (\ob^2 - \la_1)K ),
\label{es}\end{eqnarray}
where two spins are not equal as $\ms_1 = \sin^2\theta_0\ms, 
\ms_2 = \cos^2\theta_0\ms$, but have a fixed ratio $\ms_1/\ms_2 = 
\tan^2\theta_0$. They are further written by
\begin{eqnarray}
\me &=& \frac{n_1}{\pi} \frac{\sinh\rho_1}{ \sqrt{\cosh^2\rho_1 + 
\cosh^2\rho_0 - 1} }( ( \cosh^2\rho_1 + \cosh^2\rho_0 - 1)E 
- \sinh^2\rho_0K ), \nonumber \\
\ms &=& \frac{n_1}{\pi} \frac{\cosh\rho_1}{ \sqrt{\cosh^2\rho_1 + 
\cosh^2\rho_0 - 1} }( ( \cosh^2\rho_1 + \cosh^2\rho_0 - 1)E 
- \cosh^2\rho_0K ).
\end{eqnarray}
Thus the large spin limit is specified by $t \approx 1$, that is, the long
string limit $\rho_1 \rightarrow \infty$ so that $u_0 \approx u_1 
\approx 1$ and $\oa \approx \ob \gg 1$ in (\ref{kp}).
Two suitable combinations of $\me$ and $\ms$ lead to 
\begin{eqnarray}
\me - \frac{\ob}{\oa} \ms 
&=& \frac{n_1}{\pi}\frac{ \sqrt{\cosh^2\rho_1 + 
\cosh^2\rho_0 - 1} } {\sinh\rho_1}( K(t) - E(t)), \\
\me - \frac{\oa}{\ob} \ms &=& \frac{n_1}{\pi} \frac{\oa}{\sqrt{h}}K(t)
= \frac{n_1}{\pi}\frac{\sinh\rho_1}{ \sqrt{\cosh^2\rho_1 + 
\cosh^2\rho_0 - 1} }K(t).
\label{eks}\end{eqnarray}
We parametrize the large spin limit as $\eta \rightarrow 0$ for 
$\ob/\oa = 1 + \eta \; (\eta \ge 0 )$ and use $K(t) \approx 
1/2\;\ln(16/(1-t)) - (1-t)/8\;\ln(1-t) +\cdots$ with $t \approx 
1 - 2\cosh2\rho_0 \eta$ and $\ms \approx n_1/(2\pi\eta)$
for (\ref{eks}) to have 
\begin{equation}
\me - \ms + \frac{n_1}{2\pi} \approx \frac{n_1}{2\pi} \left(
\ln \frac{16\pi\ms}{n_1} - \ln\cosh2\rho_0 \right),
\label{se}\end{equation}
which includes the subleading contributions. In the large spin limit
$u_0 \approx u_1 \approx t \approx 1$ in 
$t \le u_0, t \le u_1$ which are associated
with the range (\ref{lz}), the equal angular separation $\Delta\phi$ 
(\ref{da})  for one segment turns out to be 
\begin{equation}
\Delta\phi = \frac{m}{n_1} \pi \approx \tan^{-1}\left( 
\frac{1}{\sinh2\rho_0} \right),
\label{dt}\end{equation}
where the following formula is used
\begin{equation}
\Pi(u,t) \approx \sqrt{ \frac{u}{(1-u)(u-t)} }\left( \frac{\pi}{2}
- \sin^{-1}\sqrt{ \frac{1-u}{1-t} } \right)
\end{equation}
for $u \approx t \approx 1, t \le u$. Substitution of (\ref{dt})
into (\ref{se}) yields
\begin{equation}
E - S \approx \frac{n_1\sqrt{\la}}{2\pi} \left(
\ln \frac{16\pi S}{n_1\sqrt{\la}} - 1 +  \ln\sin\frac{m\pi}{n_1} \right).
\label{en}\end{equation}
This expression for the spiky string solution with two spins is compared
with (\ref{ns}) for the  string solution of round arc shape with two equal
spins. The coefficents of the subleading 
$\ln\sin(n_2\pi/n_1)$ and $\ln\sin(m\pi/n_1)$ terms
are different where $n_2$ is the winding number in the $\theta$ direction
while $m$ is the identical winding number in both $\phi_1$ and $\phi_2$
directions. The energy-spin relation (\ref{en}) 
for the $AdS_5$ spiky string
with total spin $S$ shows the same expression as for the $AdS_3$ spiky
string with one spin $S_1$ and the winding number $m_1$
in the $\phi_1$ direction \cite{DL,MKR,JJ}
\begin{equation}
E - S_1 \approx \frac{n_1\sqrt{\la}}{2\pi} \left( \ln \frac{16\pi S_1}
{n_1\sqrt{\la}} - 1 +  \ln\sin\frac{m_1\pi}{n_1} \right).
\end{equation} 

\section{Conclusion}

By means of reduction procedure of the string sigma model to the 1-d 
Neumann integrable model \cite{AF} we have demonstrated that the string
solution \cite{LK,SRY,TT} with two equal spins in $AdS_5$ with three
angular coordinates $(\theta,\phi_1,\phi_2)$ is reproduced to be
of round arc shape in the $(\rho,\theta)$ plane, where the string
stretches in both $\rho$ and $\theta$ directions and is characterized
by the arc number and the winding number in the $\theta$ direction.
In this demonstration we have seen that the two constants of motion
$b_1, b_2$ and the three frequences $\oa, \ob = \omega_2$ are 
expressed in terms of two independent parameters, the minimum and 
maximum $AdS$ radial coordinates $\rho_0, \rho_1$.

Based on the more general reduction procedure to the integrable
Neumann-Rosochatius system \cite{ART,IK} we have constructed the spiky
string solution with two spins in $AdS_5$ by being guided from
the analysis of string equations of motion for the global
coordinates in the string sigma-model action itself.
In spite of starting with the equal choice of two frequencies
$\ob = \omega_2$ associated with the two angular directions
$\phi_1, \phi_2$ we have a spiky string solution with two unequal
spins whose ratio is fixed. We have observed that the two winding
numbers in the $ \phi_1, \phi_2$ directions shoud be the same.
We have demonstrated that the 
constant of motion $V$, the three conserved
momenta $C_a$  and the three frequences $\omega_a$ are also   
expressed in terms of two independent parameters $\rho_0, \rho_1$.
The energy and total spin
as well as two identical winding numbers are described by 
two independent parameters $\rho_0, \rho_1$ and the arc number $n_1$.
The string configuration consisting of $n_1$ arcs with a fixed 
angular separation $2\Delta\phi$  for one 
arc in the $(\rho,\phi_1)$ plane
is the same as that in the $(\rho,\phi_2)$ plane, where the spiky 
string stretches in $\rho$ as well as in both $\phi_1$ and
$\phi_2$ directions.

We have observed that the energy-spin relation in the large spin
limit for the spiky string solution with 
two spins and two identical winding numbers 
in $AdS_5$ shows the energy growing logarithmically with the
total spin and becomes the same form 
as for the spiky string with one spin 
and one winding number in $AdS_3$,
while it shows a slight difference in the coefficient of a subleading
term compared with the energy-spin relation for the string of
round arc shape with two equal spins in $AdS_5$.

\end{document}